\documentclass[manuscript,authorversion,screen,nonacm]{acmart}

\usepackage{threeparttablex}
\usepackage{multirow}

\AtBeginDocument{%
  }

\begin{document}


\title{Can metacognition predict your success in solving problems? An exploratory case study in programming}

\author{Bostjan Bubnic}
\email{b.bubnic@gmail.com}
\affiliation{%
  \institution{Faculty of Electrical Engineering and Computer Science, University of Maribor}
  \city{Maribor}
  \country{Slovenia}
}

\author{Željko Kovačević}
\affiliation{%
  \institution{Department of Computer Science and Information Technology, Zagreb University of Applied Sciences}
  \city{Zagreb}
  \country{Croatia}}
\email{zeljko.kovacevic@tvz.hr}

\author{Tomaž Kosar}
\email{tomaz.kosar@um.si}
\affiliation{%
  \institution{Faculty of Electrical Engineering and Computer Science, University of Maribor}
  \city{Maribor}
  \country{Slovenia}
}

\renewcommand{\shortauthors}{Bubnic et al.}

\begin{abstract}
 Metacognition has been recognized as an essential skill for academic success and for performance in solving problems. During learning or problem-solving, metacognitive skills facilitate a range of cognitive and affective processes, leading collectively to improved performance. 
 This study explores the predictive potential of metacognition in the second introductory programming course. 
 A two-dimensional model has been proposed, consisting of metacognitive awareness and metacognitive behavior. To evaluate the predictive capacity of metacognition empirically, an exploratory case study with 194 participants from two institutions was conducted in the second introductory programming course. A latent approach was employed to examine the associations between metacognition and performance in object-oriented programming. 
 Our findings indicate that both metacognitive dimensions have a positive effect on programming. Likewise, the results of the structural equation modeling show that 27\% of variance in programming performance is explained by metacognitive behavior. 
 Following the results, metacognition has the potential to be considered as one of the important predictors of performance in introductory programming.
\end{abstract}



\maketitle

\section{Introduction}
Metacognition is recognized as an essential construct in teaching and learning, as well as in solving problems \citep{hilton2012education}. If cognition is about doing or executing, then metacognition is about choosing and planning what to do and about monitoring what is being done \citep{Garofalo:1985}. In the past decades, aspects of metacognition have been investigated broadly in psychology and education. In their studies, researchers and educators have investigated how metacognition affects students' progression in learning, the impact of metacognitive skills on learning outcomes, and academic performance \citep[e.g.,][]{pressley2006metacognitively,schraw2006promoting,hayat2020relationships}. A substantial number of studies of metacognition have also investigated teaching strategies, approaches and methods, in order to improve students' learning \citep{zohar2013review,desoete2019metacognition}. In mathematics, previous studies have investigated the potential positive effect of metacognition on performance in solving mathematical tasks \citep[e.g.,][]{Jacobse:2012, acsik2019metacognitive,garcia2016elementary} and the potential trainability of metacognitive skills \citep[e.g.,][]{pennequin2010metacognition,desoete2003can}.

According to recent literature reviews of metacognition in computing education \citep{prather2020we,loksa2022metacognition}, the interest has surged sharply in recent years. While the earliest paper dates back to the late 1970s, there tends to be a growing interest in understanding how metacognition contributes to students' self-regulatory skills and improves learning. Furthermore, some common theories from psychology and education have recently been incorporated in the domain of Computing \citep{prather2020we}. Likewise, domain-specific theories related to metacognition in programming instruction have emerged within the discipline. This is in line with the broadly accepted notion that teaching and learning programming has some unique characteristics that could require specific metacognitive models \citep{loksa2022metacognition}. On the other hand, \citet{prather2020we} argued that empirical investigations often lack clear definition of metacognitive constructs, a well-defined research scope and limitations, and theoretical frameworks that underpin them. Regarding assessments, \citet{prather2020we} advocated that multiple measures should be incorporated into empirical investigations to triangulate effects better and increase validity (p. 11).

As a result, the primary motivation for the present study was to fill the gaps presented above. Moreover, by tracing literature reviews of metacognition in computing education \citep{prather2020we,loksa2022metacognition}, another gap has been identified. Namely, the vast majority of studies in computing education have investigated metacognition in the context of teaching and learning. Only a few studies have been found that have focused on problem-solving. Accordingly, this study attempts to shift the focus towards metacognition in solving problems. Problem-solving has been at the core of the discipline of computer science since its inception. The vast majority of problem-solving studies have examined problem-solving in relation to computer programming, targeting students in introductory programming classes \citep{tedre2018changing,deek1999software}. In this context, parallels were identified with mathematics, where problem-solving has been investigated extensively \citep{Schoenfeld:1992,Lester:2016}. As a result, the theoretical foundations and assessment instruments in this study are grounded in mathematics, where metacognition has a long history of research \citep{desoete2019metacognition}.

In this context, the objective of this exploratory case study is to evaluate the predictive potential of metacognition in programming. A two-dimensional model was used to assess the predictive capacity of metacognition. According to the results, both dimensions, metacognitive awareness and metacognitive behavior, have a positive effect on performance in object-oriented programming. Moreover, the results of structural equation modeling show that 27\% of variance in programming performance is explained by metacognitive behavior. Thus, it seems reasonable to conclude that metacognition has the potential to be used as one of the possible predictors of programming performance. 


\section{Background}
Metacognition is recognized as an essential psychological construct in teaching and learning. According to \citet{Campione:1988}, the research on metacognition stems from a paper by \citet{Tulving:1970}, in which the authors assessed the state of research on human memory critically. They emphasized that people have knowledge and beliefs about their own memory processes, and ``one of the truly unique characteristics of human memory: its knowledge of its own knowledge'' \citep[][p. 477]{Tulving:1970}. From this point on, two research paths can be observed within the literature concerning the psychological aspects of metacognition \citep{Campione:1988}. Developmental psychologists, including Flavell and Friedrichs \citep[e.g.,][]{Flavell:1970, Flavell:1979}, studied what children know about memory, and when is it that they become aware of that knowledge. This line of research has led to knowledge about cognition or \textit{metacognitive knowledge} (also referred to as metacognitive awareness). At the same time, educational and behavioral psychologists, including Brown and Campione \citep[e.g.,][]{Brown:1974,Brown:1978}, became interested in the management processes in learning, which are aimed at improving students' learning and memory skills. It was found that students' performances improved under the experimental conditions in which students were prompted explicitly to apply methods and skills by the experimenter. However, students did not apply the same skills on their own \citep{Campione:1988}. This line of research has led to explicit instruction on procedures and strategies that regulate and control human cognitive processes, also known as \textit{metacognitive regulation} (also referred to as metacognitive behavior, regulation of cognition, or metacognitive skill).

Following the initial conceptualization, numerous researchers have explored the role of metacognition in psychology \citep{Norman:2019} and education \citep{Zohar:2013,Lavi:2019}. Although the definition of metacognition has been conceptualized mostly around metacognitive knowledge and regulation of cognition, other taxonomies have also been proposed \citep[see][for a review of taxonomies]{Craig:2020,Tarricone:2011}. According to \citet{Craig:2020}, most researchers currently subscribe to the notion that metacognition is comprised of processes that monitor and increase the efficacy of cognitive procedures. Despite the consensus about the integrated nature of metacognitive components, \citet{Craig:2020} argued that the exact relation between them is yet to be established.

In the teaching and learning of mathematics, \citet{Depaepe:2010} considered metacognitive knowledge as a relatively stable information that learners possess about their own cognitive functioning, including strengths and weaknesses, as learners and problem solvers. Moreover, metacognitive knowledge refers to knowledge about where, why and how to apply certain problem-solving strategies. On the other hand, regulation of cognition is related to metacognitive skills that participants apply prior to, during, and after being engaged in learning or problem-solving \citep{Depaepe:2010}. In the broader educational literature, where the focus is more on learning than solving problems, regulation of cognition is more often considered as part of self-regulated learning. In this case, the literature is often unclear about the relationship and boundaries of metacognition, self-regulation, and self-regulated learning \citep{dinsmore2008focusing}.

According to \citet{prather2020we}, a review of the literature on metacognition and self-regulated learning in computing education shows that the earliest paper in this area dates back to 1978, with a notable increase in interest in recent years. They also found that several well-known theories from psychology and education have been applied successfully to computing. However, they noted that empirical research often lacks clear definitions of metacognitive artifacts, a defined research scope, and strong theoretical frameworks. They recommended the use of multiple assessment measures to triangulate effects and enhance validity better. \citet{loksa2022metacognition} recently expanded Prather's review to cover a broader range of computing education literature. In addition to confirm the findings by Prather and colleagues, they advocated that domain-specific theories related to metacognition in programming instruction have emerged within the discipline of computing. 

\subsection{Assessing metacognition}
Arising from the awareness of one's own thoughts and memory processes, metacognition is not directly observable, and is difficult to capture. While reviewing literature, we found several categorizations for approaches and tools to evaluate metacognition, including components and sub-components being evaluated (knowledge, regulation); capturing mechanism (self-report, interviews, performance of a task); timing (prospective: before the task, concurrent: during the task, retrospective: after the task, off-line: before or after the task, on-line: during the task); scoring (scale, coding scheme) \cite{Veenman:2019,Tarricone:2011,chen2017calibration,baars2020relation}.

Prospective and retrospective self-reports, instrumented through questionnaires and surveys, tend to be the most common way to assess metacognitive awareness in mathematical problem-solving \cite{Veenman:2019,Jacobse:2012}. In these questionnaires, respondents self-evaluate their metacognitive knowledge and strategy use, as well as frequency of its usage, through general questions and statements \citep{Jonassen:2011}. Different self-report scales, such as the Metacognitive Awareness Inventory (MAI) \citep{Schraw:1994}, Motivated Strategies for Learning Questionnaire (MSLQ) \citep{pintrich1993reliability}, and Student Thinking About Problem-Solving Scale (STAPSS) \citep{Armour-Thomas:1988} were proposed, to evaluate metacognitive knowledge and regulation, and their sub-components. The major strength of this method is the ease of administration, even on a large scale. However, \citet{Veenman:2019} reported some validity problems concerning self-report measures. Furthermore, some empirical investigations failed to show a significant correlation of metacognitive awareness and problem-solving using self-reports \citep[e.g.,][]{Depaepe:2010,Kuyper:2000}. In a recent systematic review and meta-analysis on the capacity of self-report measures, \citet{Craig:2020} argued that this method does not measure the components of metacognitive behavior adequately.

Hence, on-line measures have to be employed to capture and evaluate metacognitive behavior and metacognitive activities that occur within task processing adequately. Systematic observation, computer-logfile registration and the think-aloud protocol tend to be the most common approaches in mathematics \citep{Jacobse:2012,Veenman:2019}. According to \citet{Veenman:2019} ''thinking aloud is the only method giving access to students’ thoughts and metacognitive deliberations concurrent to task performance'' (p. 693). The major shortcoming of the think-aloud method is a time-consuming evaluation process, especially on the large scale \citep{Veenman:2019}. Researchers have also attempted to capture metacognitive behavior by incorporating self-report items into performance assessment. In particular, items assessing metacognition can be introduced before, during, or after the task performance. For instance, studies in mathematical problem-solving have integrated metacognitive judgments \citep{schraw2009measuring} and metacognitive experiences \citep{efklides2002feelings} into performance assessment \citep[e.g.,][]{Jacobse:2012,handel2020individual,tornare2015children, acsik2019metacognitive}.

\section{Conceptual model and hypothesis development}
The purpose of this exploratory case study was to evaluate the predictive potential of metacognition in object-oriented programming empirically. In line with numerous previous studies, our conceptual model considers programming as a problem-solving activity \cite[e.g.,][]{Abboud:1994,McCracken:2001,Robins:2003}. Moreover, our conceptual model relies on previous studies in mathematics, where various aspects of problem-solving have been investigated extensively \citep{Stanic:1988,Schoenfeld:1992,Lester:2016}.

In this context, our definition of metacognition aligns with a widely cited work by \citet{Garofalo:1985}, where they proposed a two-factor metacognitive taxonomy for mathematical problem-solving, consisting of: 1) knowledge of cognition, which is concerned with a person's knowledge about their cognitive abilities, processes and resources in relation to the performance within a particular problem-solving task. This factor has been sub-categorized further into personal, task-specific and strategic components. In mathematics, typical knowledge tends to be knowledge about algorithms and problem-solving strategies; 2) regulation of cognition and comprehension monitoring, which incorporates a person's awareness of which strategies to use, as well as interactions of knowledge about the problem situation with the person's strategic knowledge. Moreover, this factor is concerned with planning solution attempts, executing these plans, and monitoring the progress towards the goal state. 

Following a well-known and widely cited model of metacognition in educational psychology \citep{schraw1995metacognitive}, which has also been applied in mathematics \cite{Veenman:2019,wilson2004towards,Craig:2020}, our conceptual model defines metacognition as a two-dimensional construct. \textbf{Metacognitive awareness} is defined as an off-line measure of metacognitive knowledge, or any of its sub-components, captured prospectively or retrospectively to the problem-solving activities. \textbf{Metacognitive behavior} is defined as an on-line measure of metacognitive regulation, or any of its sub-components, captured concurrent to the problem-solving activities. 

For the purpose of this case study, metacognitive awareness and metacognitive behavior are considered as independent variables, while performance in object-oriented programming is considered as a dependent variable. The study is guided by the following hypothesis:
\begin{itemize}
\item H1: \textit{Metacognitive awareness has a positive effect on performance in object-oriented programming.}

Metacognitive awareness is concerned with a person's knowledge about their cognitive abilities and processes, including strengths and weaknesses, as learners and problem solvers. It also refers to knowledge about where, why and how to apply certain problem-solving strategies \citep{Depaepe:2010}. Mixed results have been reported regarding empirical investigations in computing education. \citet{bergin2005examining} reported a significant correlation between the application of metacognitive strategies and performance in object-oriented programming. Examining postgraduate students, \citet{yuruk2019examination} found a significant correlation between metacognition and the students' academic achievement. On the other hand, \citet{silva2023exploring} found no correlation between metacognition and programming performance. Likewise, \citet{stephenson2017exam} reported metacognitive wrappers having no effect on students' performance in a CS1 and CS2 courses. Furthermore, \citet{krause2020investigating} investigated the impact of metacognitive interventions on students' performance in CS1 course, where ``only minimal (if any) effect on students’ academic outcomes'' were identified (p. 1087).

\item H2: \textit{Metacognitive behavior has a positive effect on performance in object-oriented programming.}

Metacognitive behavior is related to metacognitive strategies that participants apply prior, during, and after being engaged in learning or solving the task \citep{Depaepe:2010}. Regarding empirical investigations in computing education, different methods have been employed to capture these skills. \citet{prather2019first} used thinking aloud in a CS1 course, to explore whether scaffolding the problem-solving process would increase metacognitive awareness. Somehow unambiguous results were obtained in their study. Namely, significant differences in metacognitive processes between control and treatment groups were only identified for students who submitted correct code in C++. No differences were demonstrated for participants who did not submit their code. Moreover, performance was not impacted, regardless of the differences in metacognition. On the other hand, two studies applying thinking aloud reported positive results. In one of the earliest studies, \citet{etelapelto1993metacognition} investigated metacognition in program comprehension between novice and expert programmers using COBOL. The results revealed that the experts were superior to the novices concerning the application of metacognitive skills. Likewise, \citet{parham2010empirical} confirmed a significant role of metacognition in computer science problem-solving. Apart from think-aloud, \citet{murphy2005computer} utilized calibration judgments, where students estimated their performance before actually taking the exam in a data structure course. According to results, students' knowledge calibrations correlated to their performance. On the contrary, \citet{lee2021targeting} assessed task-level metacognitive accuracy in the form of metacognitive judgments in the CS1 course. According to their findings, metacognition-based intervention did not resulted in significant improvements of students' metacognitive accuracy.

\end{itemize}

\section{Study design}
The present study was conducted as a part of a research project between the university in Slovenia and the university of Applied Sciences in Croatia. The primary objective of the joint research efforts has been designing and validating a new concept inventory for the second introductory programming course. The study was possible because both institutions have a common structure of introductory programming courses and C++ as a programming language. Despite this, several meetings were organized before conducting this study. Teachers from both institutions unified second introductory programming course materials and topics, discussed the teaching methods, and created a corpus of initial assessment tasks. As a result, the assessment tasks were finalized, early versions of the inventory were piloted, and the first release of the C++ concept inventory was ready to be used in the present study.

\begin{figure*}[h]
  \centering
  \includegraphics[width=\linewidth]{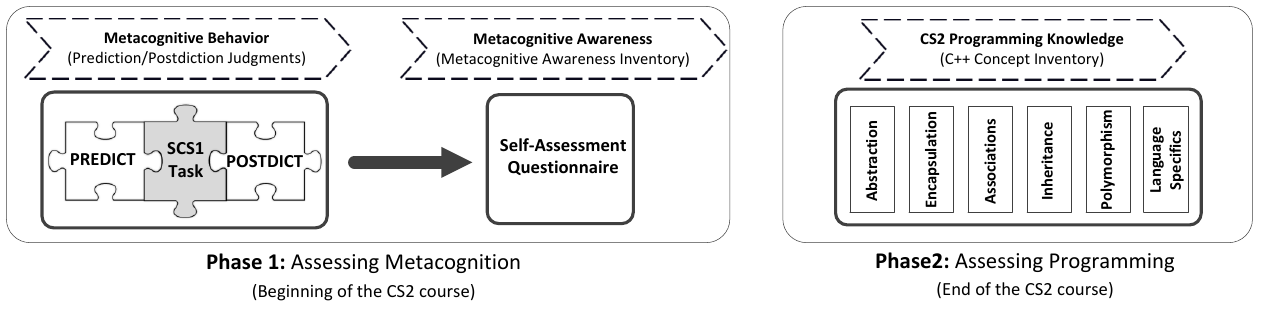}
  \caption{Experiment conduction protocol presenting both phases of the experiment}
  \label{fig:FigProtocol}
\end{figure*}

This case study was designed as a two-stage experiment. In the \textbf{first phase}, assessments of metacognitive awareness and metacognitive behavior were conducted at the beginning of the second introductory programming course. In the \textbf{second phase}, the C++ concept inventory was employed, to assess object-oriented programming performance at the end of the second introductory programming course. The experiment procedure is presented in Figure \ref{fig:FigProtocol}. 

\subsection{Participants}
The experiment included 111 students from Croatia (N = 111; 22 females; 0 non-binary; age: M = 21.7, SD = 2.8) and 83 students from Slovenia (N = 83; 6 females; 0 non-binary; age: M = 20.3, SD = 0.7) who participated in both phases of the study. Convenience sampling was used given the exploratory nature of the present study. Moreover, the objective was to include participants with diverse demographic characteristics, experiences, and backgrounds. For instance, the focus of Faculty in Croatia is on applied sciences. Therefore, many enrolled students already have at least some professional experiences. On the other hand, the Faculty in Slovenia enrolls freshmen.

The students received a course credit for their participation in both phases of the study. The credit was based on their performance. Before the study, the students were provided with a consent form explaining the purposes of the experiment.

\section{Phase 1: Assessing Metacognition}
The assessments of metacognition were conducted at the beginning of the second introductory programming course. Following our conceptual model, metacognition is considered as a two-dimensional construct, consisting of metacognitive awareness and metacognitive behavior. Metacognitive awareness is considered as an off-line measure. Therefore, the students were asked to self-assess their knowledge and regulation of cognition with the Metacognitive Awareness Inventory \citep{Schraw:1994}. Metacognitive behavior is considered as an on-line measure, and therefore it was assessed concurrent to the problem-solving tasks. In particular, the students' predictive and postdictive metacognitive judgments \citep{hacker2008metacognition} were captured during solving programming tasks. As presented in Figure \ref{fig:FigProtocol}, the metacognitive judgments were assessed first. After finishing, the students were asked to fill-in the questionnaire. This is in line with findings from \citet{Craig:2020}, who suggested that assessing metacognitive knowledge retrospectively to the problem-solving activities tends to be less biased \citep{Craig:2020}.

\subsection{Metacognitive Awareness Inventory}
The Metacognitive Awareness Inventory (MAI) \citep{Schraw:1994} was used to assess the students' metacognitive awareness. Our objective for employing this instrument was to get an off-line measure of students' metacognition. The MAI tends to be one of the most commonly used questionnaires for assessing metacognition \citep{Norman:2019}. The instrument is designed to capture the two components of metacognition, namely, knowledge of cognition and regulation of cognition \citep{Schraw:1994}. The MAI incorporates 52 items. In its original version, the response scale was implemented as ``true'' or ``false''. However, in line with the results of the instrument evaluation \citep{harrison2018evaluating}, a 5-point Likert-scale was used in our experiment (1 = Not at all typical of me, 5 = Very typical of me).

\subsection{Assessing metacognitive judgments}
The objective of assessing metacognitive judgments was to get an on-line measure of students' metacognition during their engagement in programming tasks. In this context, metacognitive judgments are used as a proxy to assess metacognitive behavior. Before solving each task, the participants had to estimate their confidence that they would solve the task successfully (prediction). Similarly, after solving each task, they had to estimate their confidence that the solution was correct (postdiction). 

From a theoretical point of view, metacognitive judgments fall under the umbrella of metacognitive monitoring and control. According to \citet{de2012improving}, metacognitive monitoring informs learners or problem solvers about necessary adaptions of their behavior, in order to circumvent obstacles encountered during learning or solving problems. The accuracy of metacognitive monitoring plays a crucial role in activating the relevant metacognitive control processes, such as planning, allocating resources, and assessing outcomes. These processes are responsible for a learner's constant progression toward learning goals or toward problem solvers' goal states \citep{bjork2013self,chen2017calibration,hacker2008metacognition}. According to \citet{efklides2001metacognitive}, problem solvers' responses to monitoring resources during problem-solving activities are reflected in metacognitive experiences. Metacognitive experiences include a broad range of experiences a person encounters during the problem-solving process, including ideas and beliefs, feelings, goals, and judgments \citep{efklides2001metacognitive}. In this context, the accuracy of metacognitive judgments can be characterized as: 1) estimates of solution correctness; 2) feelings of confidence that the solution to the problem was correct \citep{akama2006relations,Tarricone:2011}.

Regarding assessment frameworks, \citet{schraw2009measuring} described three main categories of metacognitive judgments, namely, prospective, concurrent, and retrospective. Prospective judgments require participants to make a judgment about learning or performance before performing each task or the test (predictions). Concurrent judgments require participants to make confidence or performance accuracy judgments while performing the task. For instance, if a task consists of a 15-item test, the participants are required to provide a confidence or performance accuracy judgment before and after each item. Retrospective judgments (postdictions) capture participants' evaluations of performance after completing all the test items \citep{schraw2009measuring}. More recently, \citet{handel2020individual} distinguished between local and global judgments. While local accuracy judgments are made at the task (item) level, global judgments are made at the test level. In addition, local and global judgments are also distinguished according to timing: predictions are assessed before the task and postdictions are assessed after the task \citep{handel2020individual}.

\begin{figure*}[h]
  \centering
  \includegraphics[scale=0.8]{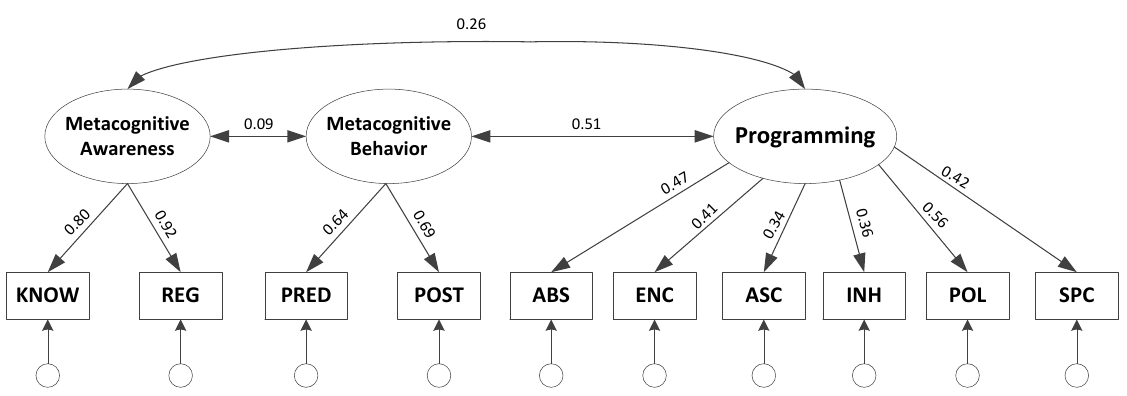}
  \caption{Standardized parameter estimates for the measurement model using confirmatory factor analysis (CFA)}
  \label{fig:figCFA}
\end{figure*}

\subsubsection{Data collection and measures} \label{secSCS1instrument}
To capture local metacognitive judgments, our instrument was grounded on previous studies that investigated metacognitive accuracy in mathematical problem-solving \citep[e.g.,][]{Jacobse:2012,garcia2016elementary,handel2020individual}. Accordingly, the following approach was incorporated into a custom web application. First, each task was presented to the students, with the instruction they had 30 seconds to read and familiarize themselves with the problem. Moreover, they were instructed explicitly not to attempt to solve the problem. After 30 seconds, the students were prompted to rate their confidence for finding the correct solution (prediction). Following the students' response, the same task was presented again, also presenting five possible answers in a multiple-choice format. After actually solving the task, the students were prompted to rate their confidence for having found the correct solution (postdiction).

Nine programming tasks from the Second CS1 Assessment (SCS1) \citep{parker2016replication} were integrated into our instrument. The SCS1 tends to be the most developed instrument for evaluating CS1 knowledge \citep{Luxton:2018}. Built in pseudocode, it was designed to assess a broad spectrum of CS1 concepts, including logical operators, conditionals, loops, functions, and arrays. These concepts were incorporated in three different types of questions: definitional, code completion, and code tracing \citep{Tew:2010}. Given that the SCS1 is used best when administered at the end of the CS1 course \citep{Parker:2021}, it was reasonable that the tasks could also serve well at the beginning of the second introductory programming course. Accordingly, three programming tasks from each question type were selected for inclusion into our instrument.

\begin{figure*}[h]
\centering
\includegraphics[scale=0.8]{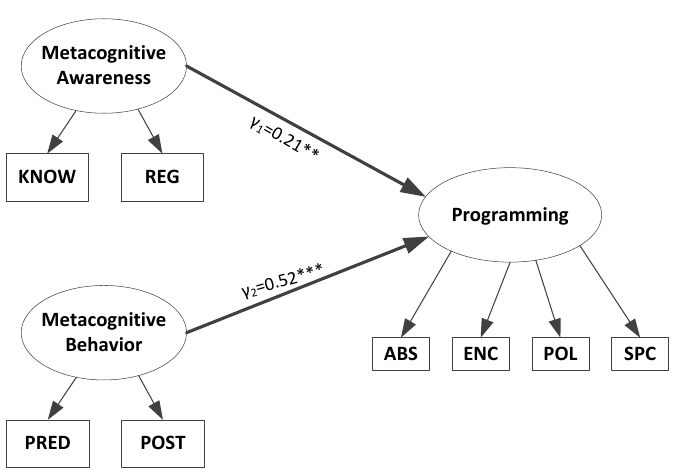}
\caption{Structural equation model for assessing H1 and H2\label{figStructural}}
\end{figure*}

Concerning the measurement scale, we followed a study by \citet{Jacobse:2012}. They incorporated task-level judgments as a part of a broader on-line measure of metacognition in mathematical problem-solving. As a result, the students' predictive (``How confident are you in solving the task correctly?'') and postdictive (``How confident are you that you selected the correct answer?'') judgments were assessed on a 3-point scale. For instance, ``I believe I can solve the task correctly'', ``I don't know whether I can solve the task'', and ``I don't believe I can solve the task correctly'' were the respective options when the students were prompted for the predictive judgment. 

We also followed a straightforward scoring scheme used previously by \citet{Jacobse:2012}. The students' prediction judgments were scored as follows. They got 1 point: 1) if they predicted they could solve the task and they actually provided a correct solution; 2) if they predicted they could not solve the task and their solution was actually wrong. Students got 0 points: 1) if their prediction was uncertain; 2) if they predicted the correct answer but provided a wrong one; 3) if they predicted the wrong answer and their solution was actually correct. Likewise, similar scoring was used for students' postdiction judgments.

\section{Phase 2: Assessing programming performance}
The objective of this phase of the study was to the assess students' object-oriented programming knowledge. The assessment was conducted at the end of the second introductory programming course.

The second introductory programming course is taught in the C++ programming language at both participating universities. The focus of the course is on learning core object-oriented programming concepts, including abstraction, encapsulation, inheritance, polymorphism, and associations. Likewise, after understanding these concepts, the students also learn details arising from the peculiarities of the C++ programming language.

\subsection{Data collection}
A C++ concept inventory was used to assess the students' object-oriented programming concepts. The new inventory was a result of a broader research initiative between the two universities participating in this study. Before being utilized in this study, the instrument had undergone three pilot studies, where the questions and respective answers were refined.

Eighteen object-oriented tasks were incorporated in the multiple-choice question format, to assess the core concepts of object-oriented programming. As presented in Figure \ref{fig:FigProtocol}, abstraction, encapsulation, inheritance, polymorphism, and associations represent the base categories. In addition, a category representing C++ language specifics was introduced. As a result, 3 questions from each category were included in the final version of the inventory. Furthermore, we also followed the question categorization introduced by \citet{Tew:2010}. Accordingly, object-oriented programming concepts were incorporated into three different types of questions: definitional, code tracing, and code completion. 


\section{Experimental results and data analysis}
A latent variable approach was used to examine the relationship between metacognitive awareness (MCA), metacognitive behavior (MCB), and performance in object-oriented programming (PROG). The Brunswik symmetry principle was followed during the data analysis \citep{wittmann1999investigating}. The principle suggests to study the relations between latent variables at the same level of aggregation or abstractness. In their study, \citet{wittmann1999investigating} demonstrated that an aggregated performance measure was best predicted by an aggregated knowledge measure. Following their approach, \textit{MCA}, \textit{MCB} and \textit{PROG} were all introduced as global measures, and were examined on the same level of aggregation.

\begin{table*}
\caption{Means, Standard Deviations, loadings, latent variables composite reliabilities (CR), and average variances extracted (AVE)\label{tbl:measurement}}
\begin{tabular*}{\textwidth}{@{\extracolsep\fill} llllllll@{} }
\toprule
Latent & Indicator & M & SD & Loadings & CR & AVE\\
\midrule
\multirow{2}{4em}{MCA}
        & KNOW & 63.101 & 10.530 & 0.797 
            & \multirow{3}{4em}{0.849} 
            & \multirow{3}{4em}{0.739} \\
        & REG & 105.649 & 20.245 & 0.919 \\\midrule
        
        \multirow{2}{4em}{MCB}
        & PRED & 9.862 & 2.338 & 0.639 
            & \multirow{3}{4em}{0.826} 
            & \multirow{3}{4em}{0.714} \\
        & POST & 10.372 & 2.856 & 0.688 \\\midrule

        \multirow{5}{4em}{PROG}
        & ABS & 1.755 & 0.892 & 0.465 \\
        & ENC & 1.563 & 0.860 & 0.409 \\
        & ASC & 1.723 & 0.912 & 0.336 
         & \multirow{3}{4em}{0.573} 
            & \multirow{3}{4em}{0.188}\\
        & INH & 2.112 & 0.861 & 0.364 \\
        & POL & 1.293 & 0.916 & 0.564 \\
        & SPC & 1.138 & 0.854 & 0.424 \\
\bottomrule
\end{tabular*}
\end{table*}

\subsection{Methodology for hypothesis testing}

Covariance-based structural equation modeling (CB-SEM) with a statistical software package, AMOS 26, was used for assessing the measurement model and for hypotheses testing \citep{blunch2012introduction}. A two-step approach was used, in line with \citet{anderson1988structural}. In the first step, a measurement model (confirmatory factor analysis) was proposed, where the \textit{MCA} was constructed as a latent variable with two manifest variables, namely, knowledge of cognition (\textit{MCA.KNOW}) and regulation of cognition (\textit{MCA.REG}). Likewise, the MCB was constructed as a latent variable with two indicators, namely, predictive judgment (\textit{MCB.PRED}) and postdictive judgment (\textit{MCB.POST}). Finally, \textit{PROG} was introduced into the model as a latent variable with six indicators, including abstraction (\textit{PROG.ABS}), encapsulation (\textit{PROG.ENC}), associations (\textit{PROG.ASC}), inheritance (\textit{PROG.INH}), polymorphism (\textit{PROG.POL}), and C++ specifics (\textit{PROG.SPEC}). In the second step, a structural equation model was proposed, its model fit was assessed, and the proposed hypotheses were tested.

\begin{table*}
\caption{Model fit indices for both the structural and measurement models}\label{tbl:indices}
\begin{threeparttable}
\begin{tabular*}{\textwidth}{@{\extracolsep\fill} lllllllllll@{} }
\toprule
 & $\chi^2$ & \textit{df} & \textit{p} ($\chi^2$) & GFI & RMR & NFI & IFI & TLI & CFI & RMSEA \\
 measurement model & 22.593 & 32 & 0.891 & 0.976 & 0.605 & 0.940 & 1.027 & 1.040 & 0.999 & 0.008 \\
 structural model & 20.059 & 18 & 0.330 & 0.973 & 1.025 & 0.942 & 0.994 & 0.990 & 0.993 & 0.025 \\
\bottomrule
\end{tabular*}
\begin{tablenotes}
\footnotesize
{\item df degrees of freedom, GFI goodness of fit index, RMR root mean square residual, NFI normed fit index, IFI incremental fit index, TLI Tucker-Lewis index, CFI comparative fit index, RMSEA root mean square error of approximation}
\end{tablenotes}
\end{threeparttable}
\end{table*}

\subsection{Measurement model} \label{secResultsMeasurement}
We used confirmatory factor analysis (CFA) to investigate the validity of the measurement model \citep{anderson1988structural, ding1995effects}. The model parameters to be estimated with their respective estimates for factor loadings and latent correlations are presented in Figure \ref{fig:figCFA}.

The results from the CFA revealed that the proposed model-data fit was satisfactory. This is reflected in the $\chi^2$ value ($\chi^2$(32) = 22.593), which is insignificant at p > 0.05, and indicates a good fit. Furthermore, additional fit indices were used, including GFI = 0.976, RMR = 0.605, NFI = 0.940, IFI = 1.027, TLI = 1.040, CFI = 0.999, and RMSEA = 0.008. GFI, NFI, IFI, TLI, and CFI were all well above 0.90. The RMSEA was lower than 0.08. In line with the representative literature \cite[e.g.,][]{maccallum1996power,hu1999cutoff}, such values also pointed towards a good fit of the model. On the other hand, RMR was higher than 0.1. According to \citet{hu1999cutoff}, RMR tends to be a positively biased measure, and that bias is greater for small sample studies. The measurement model fit indices are presented in Table \ref{tbl:indices}.

The standardized factor loadings, average extracted variance (AVE) and composite reliabilities (CR) are presented in Table \ref{tbl:measurement}. Regarding metacognitive measures, the factor loading estimates were all higher than 0.6, varying from 0.60 to 0.92. Also, AVE was above 0.7, which indicates the appropriate convergent validity of the constructs. Finally, CR were higher than the suggested threshold of 0.6, pointing toward satisfactory reliability of the metacognitive constructs. Regarding programming, the factor loadings were lower, varying from 0.34 to 0.56. Given the exploratory nature of the present study, and in line with the guidelines for identifying significant factor loadings based on simple size \citep[][p. 116]{hair2010multivariate}, items with factor loadings lower than 0.40 were removed from the model (ASC, INH). Also, CR and AVE were both lower than the suggested threshold of 0.6, which points toward lower reliability of the programming constructs.

\begin{table}
\caption{Results of the structural model}\label{tbl:structural}
\begin{threeparttable}
\begin{tabular*}{\linewidth}{@{\extracolsep{\fill}} llll }
\toprule
 H\# & Proposed relationship  & Estimate & Significance \\
\midrule
        H1 & MCA $\rightarrow$ PROG & 0.212 & p < 0.05 \\
        H2 & MCB $\rightarrow$ PROG & 0.520 & p < 0.001  \\
\bottomrule
\end{tabular*}
\begin{tablenotes}
\footnotesize
{\item MCA metacognitive awareness, PROG performance in object-oriented programming, MCB metacognitive behavior}
\end{tablenotes}
\end{threeparttable}
\end{table}

\subsection{Results of the structural model} \label{secResultsStructural}
A structural equation model was used to test our hypotheses. Structural paths were proposed from MCA to PROG and from MCB to PROG. In line with results of the confirmatory factor analysis, two indicators (ASC, INH) were not included in the structural model. The structural equation model is presented in Figure \ref{figStructural}.

The overall model fit was assessed before analyzing the structural paths. As presented in Table \ref{tbl:indices}, $\chi^2$ was again non-significant ($\chi^2$(18) = 20.059, \textit{p > 0.05}). Likewise, the other fit indices, with the exception of RMR, which is sensitive to lower sample sizes, were inside the suggested intervals, pointing toward a satisfactory model fit.
The results of the structural path analysis are presented in Table \ref{tbl:structural}. The findings showed that MCA had a positive and significant impact on PROG ($\gamma_{1}$ = 0.21, \textit{p} < 0.05). Hence, we can support our H1. The findings also indicated that MCB had a strong positive impact on PROG ($\gamma_{2}$ = 0.52, \textit{p} < 0.001), which confirmed our H2.

\section{Discussion}
This exploratory study investigates whether metacognition could potentially be used as a predictor of performance in the second introductory programming course. A two-dimensional model, consisting of metacognitive awareness and metacognitive behavior, was used to examine the associations between metacognition and performance in object-oriented programming. Metacognitive awareness was introduced as an off-line measure, and was assessed retrospective to the students' performance in programming. Metacognitive behavior was used as an on-line measure, which was assessed concurrent to the students' programming tasks. We found that both metacognitive dimensions had a positive effect on programming. Following our results, metacognition has the potential to be considered as a predictor of performance in introductory programming.

Although metacognition has a long history of investigations in programming and computer science, the present study may be the first that applied a multi-method approach. Regarding the comparison of the two metacognitive dimensions, our results indicate that the on-line measure had a stronger effect than the off-line measure. This is in line with previous results of multi-method approaches to assess metacognition in psychology and education. For instance, \citet{Craig:2020} conducted a systematic literature review and meta-analysis to evaluate the ability of self-report measures to measure all aspects of metacognition adequately. According to their results, self-reports did not measure components of metacognitive behavior adequately. Likewise, \citet{Veenman:2019} compared off-line and on-line measures of metacognition in mathematics. Their findings suggest that on-line instruments should be the preferred method for assessing metacognitive skillfulness.

Beyond the predictive potential, this study attempts to bring attention to problem-solving aspects of metacognition. By examining literature reviews in computing education, we found that most studies have examined metacognition within the teaching and learning context, while only a limited number have focused on problem-solving. Historically, problem-solving has been a foundational aspect of programming and computer science. In this context, similarities were drawn with mathematics, a discipline where cognitive, metacognitive and affective aspects of problem-solving have been investigated thoroughly. As a result, our study advances the field of Metacognitive research in computing education by drawing on theoretical foundations and assessment methodologies from mathematics, where metacognition has a well-established research history.

Concerning instruments' timing, metacognitive awareness was assessed retrospective to the students' performing computer programming tasks (see Figure \ref{fig:FigProtocol}). This sequence of instruments was chosen according to suggestions by \citet{Craig:2020}. Namely, the authors argued that concurrent and retrospective metacognitive measurements give more reliable results then prospective and concurrent \citep{Craig:2020}.

Regarding our first hypothesis, our results revealed a positive and significant correlation between metacognitive awareness and students' performance in object-oriented programming. This outcome was in line with what we anticipated, and was also supported by findings from prior studies \citep[e.g.,][]{bergin2005examining,yuruk2019examination}. Metacognitive awareness was introduced as an off-line measure, which was assessed with the Metacognitive Awareness Inventory (MAI). By leveraging the MAI, our study advances the field as follows. First, by reviewing the two recent literature reviews of metacognition in computing education \citep{prather2020we,loksa2022metacognition}, we found that the vast amount of previous empirical investigations utilized the Motivated Strategies for Learning Questionnaire (MSLQ). In this context, \citet{silva2023exploring} cited some prior studies in programming that reported negative or conflicting results with the MSLQ. Moreover, according to the results of their study, authors called for further investigations regarding programming performance and self-report measures \citep{silva2023exploring}. Second, as the MAI tends to be the most commonly used instrument in problem-solving research \citep{Jonassen:2011}, our approach shifted the emphasis from teaching and learning toward performance in solving problems. 

Regarding our second hypothesis, our results showed that metacognitive awareness has a strong positive effect on the students' performance in object-oriented programming. Likewise, the results of the structural equation modeling showed that 27\% ($R^{2}$ = 0.27) of variance in programming performance was explained by metacognitive behavior. We anticipated metacognitive behavior having a greater effect on programming performance than metacognitive awareness. Likewise, some previous studies identified strong effects of metacognitive behavior on problem-solving performance in mathematics \citep{desoete2008multi,Veenman:2019}. Task-level metacognitive accuracy judgments were used to assess metacognitive behavior as an on-line measure of metacognition. This method has been applied previously in mathematical problem-solving \citep{garcia2016elementary,Jacobse:2012} as well as across disciplines \citep{handel2020individual}. In this context, this study contributes to metacognitive awareness research in programming, traditionally dominated by think-aloud as the on-line measure.

\section{Threats to validity}
This section presents potential threats to construct, internal, and external validity in our experimental study.

\subsection{Construct validity}
The primary concern of construct validity in this experiment is related to the validity of the employed instruments. In the first phase, we used a custom instrument for assessing metacognitive behavior, which has not undergone validations. To mitigate the situation somehow, we implemented the following. First, metacognitive awareness was introduced as a second measure of metacognition. It was evaluated with the Metacognitive Awareness Inventory, which underwent several independent evaluations. Second, as described in Section \ref{secSCS1instrument}, our metacognitive instrument incorporated questions from the SCS1 concept inventory. Tasks from this inventory have been the subject of different independent validations \citep[e.g.,][]{xie2019item}. As a result, our first selection criterion was the reliability of the questions from each category, for inclusion into our instrument.

In the second phase, our newly developed C++ concept inventory was employed, to evaluate the students' object-oriented programming performance. The following measures were taken to mitigate validity. Three pilot studies were conducted before this study. According to the results, the reliability of the assessment items was satisfactory. However, all the pilot studies were conducted with participants who already passed the second introductory programming course. In this context, there was a concern that tasks could be too difficult for the second introductory programming students. To balance this concern, the questions from the first phase of the experiment were also selected according to difficulty criteria. Namely, when selecting questions for inclusion in the metacognitive behavior instrument (see the previous paragraph), our second inclusion criteria was difficulty. Accordingly, questions of higher difficulty were selected from the SCS1 assessment. Finally, due to low factor loadings, the results from inheritance (INH) and polymorphism (POL) were excluded from the analysis (see also Sections \ref{secResultsMeasurement} and \ref{secResultsStructural}). Despite the described procedures, the factor loadings and average variances extracted for the C++ concept inventory were lower than anticipated (see Table \ref{tbl:measurement}).

\subsection{Internal validity}
Potential internal validity threats exist because of the sample size and convenience sampling. Moreover, due to the lower than expected students' participation, data from two different samples were combined for the purpose of structural equation modeling. However, considering the exploratory nature of the study, and given that the variables were only examined as global measures, the potential level of implications is acceptable.

Another internal validity concern is associated with potential student drop outs between both phases of the experiment (mortality) and because of missing values. In terms of mortality, students who completed both phases of the experiment may exhibit higher motivation. With respect to missing data, fourteen students did not finish either the first or second phase of the experiment. They were all excluded from the analysis, which introduced a selection bias. Given that the missing values constituted less than 8\% of the total participants, the potential impact was considered acceptable.

\subsection{External validity}
The results are limited primarily to the second introductory programming course. The results' generalizability was also constrained, due to the experiment's specific context. First, the study relied on a convenience sample. Second, demographic and cultural factors might have influenced the outcomes. To mitigate these concerns, students from two institutions in different countries participated in the study.

An additional concern regarding external validity relates to the latent variable \textit{PROG}, which exhibited an average variance extracted (AVE) value well below the recommended threshold of 0.5 (AVE = 0.188). When the AVE is below this benchmark, the measurement model's convergent validity may be weaker than anticipated.

Finally, exploratory studies require replications in the controlled multi-institutional (and multi-national) environment, to confirm the level of confidence before drawing general conclusions.

\section{Conclusions and future directions}
This case study explored the predictive potential of metacognition in second introductory programming. A multi-method approach was used to assess metacognition, while our newly developed C++ concept inventory was utilized to assess object-oriented programming performance. According to the results, metacognition has the potential to be considered as a predictor of performance. Despite the limitations inherent in the exploratory nature of this study, we believe that it contributes to the segments of metacognitive research where previous investigations have been lacking. In this regard, we encourage the computing education community to replicate our study, and re-evaluate the level of confidence suggested by our results.

Regarding future work, we would like to investigate the on-line measures of metacognition further. In particular, we would like to explore metacognitive behavior in terms of metacognitive processes and respective sub-processes while an individual is performing problem-solving tasks in programming. On the other hand, the ChatGPT era is shifting the focus from cognitive to other than cognitive aspects of problem-solving in programming. In this context, it would be interesting to explore the socio-psychological aspects of metacognition, such as the Dunning–Kruger effect.

\bibliographystyle{ACM-Reference-Format}
\bibliography{MyPaper}


\end{document}